\documentclass[%
reprint,
superscriptaddress,
twocolumn,
amsmath,amssymb,
aps,
prb,
longbibliography,
floatfix
]{revtex4-2}

\usepackage{graphicx}
\usepackage{dcolumn}
\usepackage{bm}
\usepackage{amsmath}
\usepackage{amssymb}
\usepackage{multirow} 
\usepackage{color}
\usepackage[T1]{fontenc}
\usepackage[colorlinks=true,linkcolor=blue,citecolor=blue]{hyperref}
\usepackage{booktabs}
\usepackage{tabularx}
\usepackage[english]{babel}
\usepackage[utf8]{inputenc}
\usepackage{enumerate}
\usepackage[cmyk,dvipsnames]{xcolor}

\usepackage[normalem]{ulem}

\definecolor{darkyellow}{rgb}{0.6, 0.6, 0.0}

\definecolor{pacificb}{HTML}{1CA9C9}

\date{ \today}

\date{ \today}

\begin{document}

\title{Optimal field-free magnetization switching via spin-orbit torque on the surface of a topological insulator}

\author{Ivan P. Miranda}
\thanks{These two authors contributed equally to this work.}
\affiliation{Department of Physics and Electrical Engineering, Linnaeus University, SE-39231 Kalmar, Sweden}
\affiliation{Department of Physics and Astronomy, Uppsala University, SE-75120 Uppsala, Sweden}

\author{Grzegorz J. Kwiatkowski}
\thanks{These two authors contributed equally to this work.}
\affiliation{Science Institute, University of Iceland, 107 Reykjav\'ik, Iceland}

\author{Cecilia M. Holmqvist}
\affiliation{Department of Physics and Electrical Engineering, Linnaeus University, SE-39231 Kalmar, Sweden}

\author{Carlo M. Canali}
\affiliation{Department of Physics and Electrical Engineering, Linnaeus University, SE-39231 Kalmar, Sweden}

\author{Igor S. Lobanov}
\affiliation{Department of Physics, ITMO University, 197101 St. Petersburg, Russia}

\author{Valery M. Uzdin}
\affiliation{Department of Physics, ITMO University, 197101 St. Petersburg, Russia}

\author{Andrei Manolescu}
\affiliation{Department of Engineering, Reykjav\'ik University, 102 Reykjav\'ik, Iceland}

\author{Pavel F. Bessarab}
\email[Corresponding author: ]{pavel.bessarab@lnu.se}
\affiliation{Science Institute, University of Iceland, 107 Reykjav\'ik, Iceland}
\affiliation{Department of Physics and Electrical Engineering, Linnaeus University, SE-39231 Kalmar, Sweden}

\author{Sigurdur  I. Erlingsson}
\affiliation{Department of Engineering, Reykjav\'ik University, 102 Reykjav\'ik, Iceland}

\begin{abstract}
We present an optimal field-free protocol for current-induced switching of a perpendicularly magnetized ferromagnetic insulator nanoelement on the surface of a topological insulator. The time dependence of in-plane components of the surface current, which drives the magnetization reversal via the Dirac spin-orbit torque with minimal Joule heating, is derived analytically as a function of the switching time and material properties. Our analysis identifies that energy-efficient switching is achieved for vanishing damping-like torque. The optimal reversal time that balances switching speed and energy efficiency is determined. When we compare topological insulators to heavy-metal systems, we find similar switching costs for the optimal ratio between the spin-orbit torque coefficients. However, topological insulators offer the advantage of tunable material properties.  
Finally, we propose a robust and efficient simplified switching protocol using a down-chirped rotating current pulse, tailored to realistic ferromagnetic/topological insulator systems.

\end{abstract}

\maketitle

\section{Introduction}   

Spin-orbit torque (SOT), a torque based on the conversion of charge current to spin current and interface spin accumulation via spin-orbit interaction, provides an efficient mechanism for electrical control of magnetization~\cite{gambardella2011, miron2011, liu2012, manchon2019}. Magnetization switching is an important application of the SOT~\cite{aradhya2016, garello2014} as it can be assigned to represent logical operations in nonvolatile information technologies. The challenge is to minimize the energy cost of the switching. An SOT-induced magnetization reversal can be realized by applying an in-plane current in a heavy-metal (HM) layer on which a switchable ferromagnetic (FM) element is placed. In this case, the SOT can originate from the inverse spin galvanic effect~\cite{edelstein1990,belkov2008,manchon2008} and the spin Hall effect~\cite{dyakonov1971,hirsch1999,ando2008,liu2011,sinova2015} in the HM. Both mechanisms of such HM SOT give rise to fieldlike (FL) and dampinglike (DL) torques, whose geometrical forms coincide with those of the spin-transfer torque~\cite{manchon2019}. Conventional switching protocols rely solely on the DL SOT and utilize a unidirectional current~\cite{lee2013,fukami2016}, inherently requiring the application of an external magnetic field to achieve deterministic magnetization reversal \cite{miron2011,liu2012,shao2021}. However, switching with much lower current densities can be achieved by using both in-plane components of the current~\cite{zhang2018}, also allowing for a field-free operation.

Analysis based on the optimal control theory has previously revealed a theoretical limit for the minimum energy cost of the SOT-induced magnetization reversal in the FM/HM system, identified the corresponding optimal switching current pulse as a function of the reversal time and relevant material properties, and uncovered 
a sweet-spot ratio of the FL and DL torques which allows one to achieve a particularly efficient switching by a down-chirped rotating current pulse~\cite{vlasov2022}. 
The optimization of magnetization switching protocols is both fundamentally interesting and technologically important, as it enables the identification of the right combination of material properties for energy-efficient applications. 
\\

Topological insulators (TIs), i.e., materials characterized by an insulating energy gap in the bulk and gapless edge or surface states protected by time-reversal symmetry~\cite{hasan2010, XLQi2011, Ando2013, Bansil2016}, represent a promising alternative to HM systems in the context of electrical control of magnetization~\cite{Mellnik2014}. The spin-momentum locking of the Dirac electrons at the surface of a TI results in a large charge-spin conversion, thereby enabling a significant SOT on an adjacent FM \cite{Mellnik2014}, hereafter referred to as Dirac or TI SOT.
Remarkably, such SOT has a different geometrical form compared to the one found in HM systems~\cite{ndiaye17:014408}, affecting the magnetization dynamics \cite{Nagaosa2024}. It is therefore interesting to explore how much the Dirac SOT-induced magnetization dynamics can be optimized and whether using TI systems offers benefits for achieving energy-efficient magnetization switching.

In this article, we present a complete analytical solution to the problem of energy-efficient magnetization switching in an FM insulator nanoelement with perpendicular magnetic anisotropy (PMA) by means of Dirac SOT realized in TI systems. We show that the optimal switching protocol, i.e. the protocol that minimizes the energy cost associated with Joule heating, involves both in-plane components of the surface current whose time dependence is determined by the materials' properties and the switching time. In this optimal scenario, the magnetization reversal is deterministic and does not require any external magnetic field. We obtain noteworthy exact dependencies concerning the optimal switching and compare them with the results obtained earlier for the HM systems~\cite{vlasov2022}. We demonstrate that the best scenario for magnetization switching is realized for vanishing DL Dirac SOT. For realistically small DL SOT, the optimal switching is achieved in a particularly simple protocol involving a down-chirped rotating current, which can be readily generated in experimental setups. 
Finally, we discuss various possibilities to tune the SOT coefficients to achieve the most efficient switching. We conclude that TI systems offer a convenient and robust platform for energy-efficient magnetization control, providing several advantages over other systems. 

\begin{figure}[!ht]
\centering
\includegraphics[width=\columnwidth]{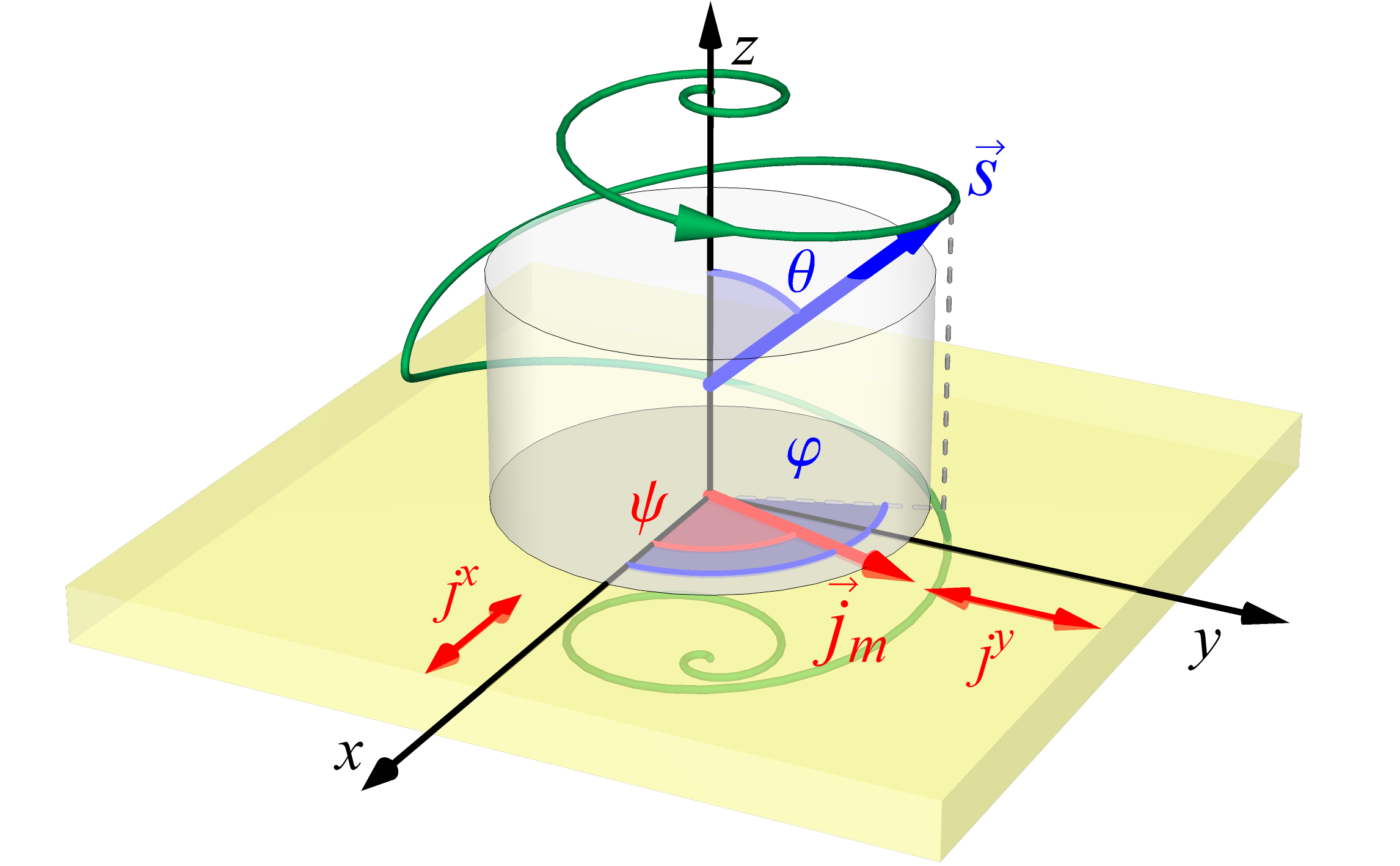}
\caption{\label{fig:1} Energy-efficient switching of a PMA nanoelement (cylinder) by a Dirac spin-orbit torque induced by an optimal 2D electric current pulse $\vec{j}_m$ flowing at the surface of a topological insulator substrate. The calculated optimal control path of the switching for $\alpha=0.1$ and $\xi_D= \xi_F$ and $T=50\tau_0$
is shown with the green line. 
The direction of the normalized magnetic moment $\vec{s}$ of the element (optimal current $\vec{j}_m$) is shown with the blue (red) arrow.}
\end{figure}

\section{Model}
The simulated PMA nanoelement placed on the surface of a TI is illustrated in Fig.~\ref{fig:1}. The magnetic moment of the nanoelement is reversed by a surface current via Dirac SOT. The nanoelement is assumed to be small enough that its magnetization remains essentially uniform during the reversal process (single-domain) \footnote{For free nanocrystalline Y$_3$Fe$_5$O$_{12}$ (YIG) particles, for example, the critical size is about 190 nm \cite{Sanchez2002}.}, but not so small that  quantum correlations become significant \cite{Liu2024}, allowing the classical spin dynamics to remain a valid phenomenological model. We also assume the nanoelement to be a ferromagnetic insulator so as to ensure that the current flows mostly via the TI host \cite{Mellnik2014}. The energy $E$ of the system is defined by the orientation of the nanoelement's magnetic moment with respect to the anisotropy axis, which is perpendicular to the surface of the TI (see the reference frame in Fig.~\ref{fig:1}), 

\begin{equation}
    \label{eq:energy}
    E = -Ks_z^2,
\end{equation}

\noindent where $s_z$ is the out-of-plane component of the normalized magnetic moment $\vec{s}$ and $K>0$ is the anisotropy constant. 
We aim to identify the optimal current pulse that reverses the magnetic moment from $s_z=1$ at $t=0$ to $s_z=-1$ at $t=T$, with $T$ being the switching time. 
Both the amplitude and the direction of the current $\vec{j}$ on the surface of the TI are allowed to vary in time:

\begin{equation}
\label{eq:current-j}
\vec{j}(t)=j(t)\left[\cos\psi(t), \sin\psi(t),0\right].
\end{equation}

The efficiency of reversal is naturally defined by the amount of Joule heating generated in the resistive circuit during the switching process~\cite{tretiakov_2010}. In particular, the optimal reversal is achieved when the cost functional 
\begin{equation}
\label{eq:cost}
    \Phi = \int_0^T |\vec{j}|^2dt,
\end{equation}
is minimized \footnote{The surface resistivity $\rho$ of the TI is assumed to be uniform and constant over time, which is a reasonable assumption, particularly for thin TI films~\cite{Wang2017}.}. This optimal control problem is subject to a constraint imposed by the zero-temperature Landau-Lifshitz-Gilbert (LLG) equation describing the dynamics of the magnetic moment induced by the Dirac SOT~\cite{ndiaye17:014408}: 

\begin{eqnarray}
\label{eq:eom}
    \dot{\vec{s}} = &-&\gamma\vec{s}\times\vec{b}
	+ \alpha\vec{s}\times\dot{\vec{s}} \nonumber\\
	&+& \gamma\xi_{F}\vec{s}\times(\vec{j}\times\hat{e}_z) 
	+ \gamma\xi_{D}s_z(\vec{s}\times \vec{j}).
\end{eqnarray}

Here, $\gamma$ is the gyromagnetic ratio, $\alpha$ is the damping parameter, $\hat{e}_z$ is the unit vector along the $z$ axis, and $\vec{b}$ is the anisotropy field: $\vec{b} \equiv -\mu^{-1} \partial E / \partial \vec{s}$, with $\mu$ being the magnitude of the magnetic moment. The third and the fourth terms on the right-hand side of Eq.~(\ref{eq:eom}) represent the FL and DL components of the SOT, respectively, where we note a fundamental difference in the geometrical form of the DL component from that realized in FM/HM systems ($\propto\vec{s}\times[\vec{s}\times(\vec{j}\times\hat{e}_z)]$) \cite{manchon2019}. The coefficients $\xi_F$ and $\xi_D$ can be written in terms of microscopic parameters of the system \cite{ndiaye17:014408,manchon2019}. Both $\xi_D$ and $\xi_F$ are assumed to be independent of $\vec{s}$. The validity of this assumption is discussed in Section \ref{sec:tunability}.  
Moreover, the effect of the Oersted field generated by $\vec{j}$ is neglected in Eq. (\ref{eq:eom}) as it is by 3-4 orders of magnitude smaller than the typical anisotropy field \footnote{Assuming a charge current flowing through an infinite plane (i.e., across the surface of a TI) and applying the quasi-static approximation, the time-dependent Oersted field generated by the current density $\vec{j}$ lies parallel to the plane. Its magnitude is given by $B_0=\mu_0 j/2$ with $\mu_0$ being the magnetic constant. Using realistic experimental values from Ref. \cite{Mellnik2014}, we estimate $B_0\approx(2.5-8)\times10^{-6}$ T. A similar estimate ($B_0\approx3\times10^{-5}$ T) can be obtained with the setup used by Wang \textit{et al.} \cite{Wang2017}. The maximum value of the anisotropy field, $b_\text{max}=2K/\mu$, can be estimated using \textit{ab-initio} and experimental results for the magnetic moment~\cite{Gorbatov2021} and anisotropy~\cite{Wang2014, Fu2017} of YIG. This yields $b_\text{max}\approx10^{-2}$ T, which is 3 to 4 orders of magnitude larger than $B_0$.}. 

In what follows, 
we focus on how the ratio of the SOT coefficients influences the switching cost. Using the approach from our previous study~\cite{vlasov2022},  we introduce the following parametrization:

\begin{equation}
\label{eq:TIcomponents}
    \xi_F = \xi \cos\beta, \qquad \xi_D = \xi \sin\beta,
\end{equation}
where $\xi = \sqrt{\xi_F^2+\xi_D^2}$ is the magnitude of spin-orbit torque and $\beta$ is a dimensionless parameter characterizing the proportion between its components.

\section{Results}
To find the optimal switching current $\vec{j}_m(t)$ \footnote{Throughout this article, quantities indexed with $m$ represent optimal values corresponding to the minimum energy cost of switching.}, we first express $\Phi$ in terms of the switching trajectory and then minimize it so as to find the optimal control path (OCP) $\vec{s}_m(t)$ for the switching process; after that, $\vec{j}_m(t)$ is derived from the OCP. A similar procedure was applied previously in the context of magnetization switching induced by an applied magnetic field~\cite{kwiatkowski2021optimal} and SOTs in HM systems~\cite{vlasov2022}.  Here, we briefly report the main expressions, but a complete analytical solution can be found in the Supplemental Material (SM).

\subsection{Optimal magnetization switching}
\label{sec:optimal-switching}
We first notice that the direction $\psi$ of the current [see Eq.~(\ref{eq:current-j})] can be written as a function of the spherical coordinates $\theta$ and $\varphi$ of the magnetic moment $\vec{s}$ (see Fig.\ \ref{fig:1} and SM): 
\begin{equation}
\label{eq:current_direction}
    \psi = \varphi - \arctan\left[\frac{\cos\theta\left(\alpha\cos\beta + \sin\beta\right)}{\cos\beta -\alpha \cos^2\theta \sin\beta}\right].
\end{equation}
 
It follows from Eqs. (\ref{eq:eom}) and (\ref{eq:current_direction}) that the time-dependent amplitude of the current can 
be expressed in terms of $\theta$ and $\dot{\theta}$ as 

\begin{equation}
\label{eq:current_amplitude_final}
    j = j_0\frac{\left(1+\alpha^2\right)\tau_0 \dot{\theta}+\alpha\sin\theta \cos\theta}{\sqrt{(1+\alpha^2\cos^2\theta)(1-\sin^2\theta\sin^2\beta)}},
\end{equation}
where $j_0=K(\mu\xi)^{-1}$ and $\tau_0=\mu \left(2 K \gamma\right)^{-1}$. Equation~(\ref{eq:current_amplitude_final}) indicates that the current amplitude is the lowest for $\beta=0$ and $\beta=\pi$ leading to the most energy efficient switching. On the other hand, for $\beta = \pi/2$ and $\beta = 3\pi/2$ the switching is impossible as the current amplitude is infinite at the top of the energy barrier ($\theta = \pi/2$). This is a consequence of the geometrical form of the Dirac SOT, for which the DL torque vanishes for an in-plane orientation of the magnetic moment. 
Notably, the current amplitude is independent of the azimuthal angle $\varphi$. As a result, the energy cost of switching, $\Phi$, is a functional of the polar angle $\theta$ alone, which simplifies the minimization. 

We are interested in minimizing the cost functional $\Phi$ defined in Eq. (\ref{eq:cost}) with the boundary conditions of $\theta(0)=0$, $\theta(T)=\pi$. The solution to the corresponding Euler-Lagrange equation for $\theta(t)$ is expressed in terms of Jacobi elliptic functions \cite{abramowitz1948handbook}. Subsequently, $j_m(t)$ is obtained from Eq. (\ref{eq:current_amplitude_final}). Subsequently, $\varphi_m(t)$ is calculated via direct integration of the equation of motion and $\psi_m(t)$ is obtained using Eq.~(\ref{eq:current_direction}). Regardless of the system parameters, the OCP always crosses the energy barrier $\theta = \pi/2$ at $t=T/2$ which is a result of a more general symmetry $\theta_m\left(T/2+t'\right)=\pi-\theta_m\left(T/2-t'\right)$, for $0\le t^\prime\le T/2$.

According to Eq. (\ref{eq:current_direction}), the current rotates following the magnetic moment as it precesses around the anisotropy axis, gradually changing its frequency. The exact expressions for the frequency and amplitude are quite complex (see the SM for details), but an analytical expression for the average current amplitude has a relatively simple form: 
\begin{equation}\label{eq:average_current}
    \langle j_m\rangle = \frac{2\left(1+\alpha^2\right)\mathcal{K}\left[\sin^2\beta+\alpha^2\cos^2\beta\right]}{\xi\gamma T},
\end{equation}
where $\mathcal{K}[k]=\int_{0}^{\pi/2} dr\left(1-k\sin^2r\right)^{-1/2}$ is the complete elliptic integral of the first kind. Note that the average current is independent of $K$. Similar to our earlier findings on the average current~\cite{vlasov2022} and magnetic field~\cite{kwiatkowski2021optimal}, the independence of $\langle j_m \rangle$ from $K$ arises from the cancellation of the anisotropy field's relaxation contribution during the switching process. Specifically, this field opposes switching for $\theta < \pi/2$ but facilitates reaching the final state for $\theta > \pi/2$, leading to a net null effect.
However, the anisotropy \textit{does} affect the amplitude of the optimal current: for 
large $K$, $j_m(t)$ localizes in time, sharp (localized in time) and large pulse, 
while for small $K$, it becomes time independent. 

For $\alpha=0$, Eq.\ (\ref{eq:average_current}) matches the average current $\langle j_m^{\textnormal{HM}}\rangle$ obtained in Ref. \cite{vlasov2022} for FM/HM systems. Assuming the same values of $\xi$ and $T$ for both FM/HM and FM/TI systems, 
 the ratio $\langle j_m\rangle/\langle j_m^{\textnormal{HM}}\rangle$ stays close to unity for $\alpha<10^{-2}$ and for $\beta$ values in the range $[0,\frac{\pi}{2})$. However, for 
$\alpha>10^{-2}$, $\langle j_m\rangle$ can become much smaller or much larger than $\langle j_m^{\textnormal{HM}}\rangle$, depending on the value of $\beta$. The analytical expression for $\langle j_m\rangle/\langle j_m^{\textnormal{HM}}\rangle$ providing its dependency on $\alpha$ and $\beta$ is given in the SM.

\begin{figure*}[t]
\centering
\includegraphics[width=0.65\textwidth]{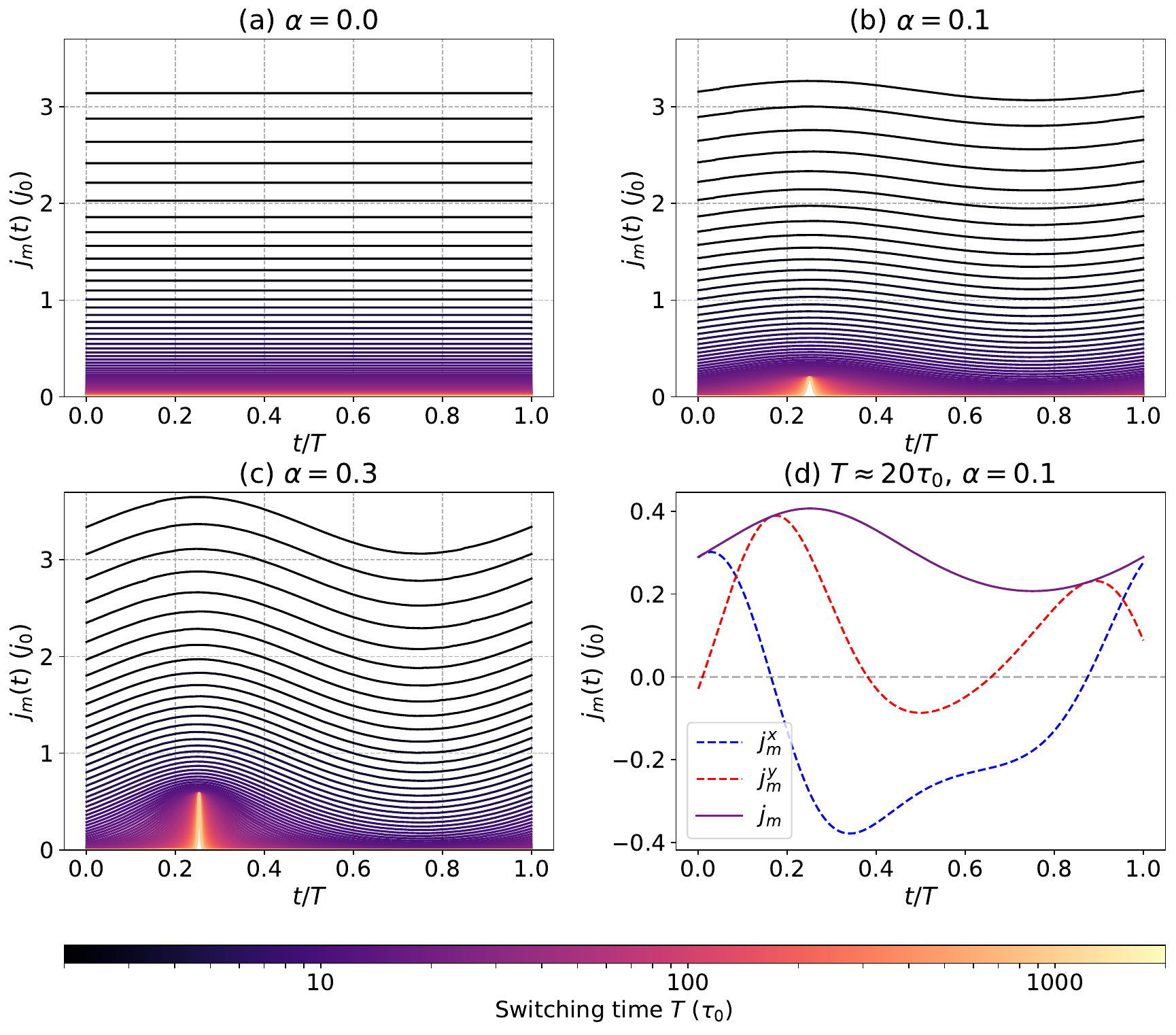}  
\vspace{-4mm}
\caption{(Color online) (a-c) Calculated magnitude of the optimal switching current as a function of time for several values of the damping parameter and switching time, as indicated in the legend; (d) Calculated magnitude and in-plane components of the optimal switching current as functions of time for $T=20\tau_0$ and $\alpha=0.1$. In all cases $\beta=0$, which corresponds to the lowest energy cost of switching.
}
\label{fig:evolution-optimal-current}
\end{figure*}

Figure \ref{fig:evolution-optimal-current} shows the $j_m(t)$ profiles 
as functions of the reduced time $t/T$ for various values of the switching time $T$, considering vanishing DL torque ($\beta=0$). As $\alpha$ increases, the difference between the maximum and minimum values of $j_m$ becomes more pronounced. These extremal values of $j_m$ occur at specific points in time: the maximum at $t/T = 1/4$ and the minimum at $t/T = 3/4$. For zero damping, the current amplitude is time independent. 

\subsection{Minimum energy cost of magnetization switching}
\label{sec:minimum-energy}

As in the previously studied cases \cite{kwiatkowski2021optimal, vlasov2022}, $\Phi_m$ monotonically decreases with $T$ and demonstrates two distinct asymptotic behaviors. For short switching times, $T\ll \left(\alpha+1/\alpha\right)\tau_0$, the contribution from the internal dynamics driven by the anisotropy becomes negligible and the solution approaches that of a free magnetic moment, resulting in
\begin{equation}
    \Phi_{0} = \frac{4(1+\alpha^2)^2\mathcal{K}^2[1-(1+\alpha^2)\cos^2\beta]}{\xi^2 \gamma^2 T}.
\end{equation}
On the other hand, for $T\rightarrow \infty$ the energy cost approaches a lower limit of
\begin{equation}
    \Phi_\infty = \frac{4\alpha K (1+\alpha^2) \ln\left[(1+\alpha^2)\cos^2(\beta)\right]}{\gamma\mu\xi^2\left[(1+\alpha^2)\cos^2(\beta)-1\right]}.
\end{equation}
The intersection of these asymptotics gives an optimal switching time $T_{\textnormal{opt}}$ for which a balance between switching speed and energy efficiency is achieved. When $\beta=0$, $T_{\textnormal{opt}}$ coincides with that found for FM/HM systems \cite{vlasov2022}, which has the simple approximate form

\begin{equation}
\label{eq:approximate-topt}
T_{\textnormal{opt}}=\frac{(1+\alpha^2)\pi^2}{2\alpha}\tau_0,
\end{equation}

\noindent correct up to $\approx0.5\%$ for $0\leq\alpha\leq1$. Note that Eq. (\ref{eq:approximate-topt}) implies that $T_{\textnormal{opt}}\geq\pi^2\tau_0\approx10\tau_0$ when $\alpha=1$. For arbitrary values of $\beta$, $T_{\textnormal{opt}}$ can differ significantly from that obtained for the FM/HM case, especially if $\alpha\rightarrow1$ and/or $\beta\rightarrow\pi/2$. A complete comparison between the optimal switching times for both Dirac and HM SOT, considering various values of $\alpha$ and $\beta$, can be found in the SM.

\begin{figure}[h]
\centering
\includegraphics[width=220pt]{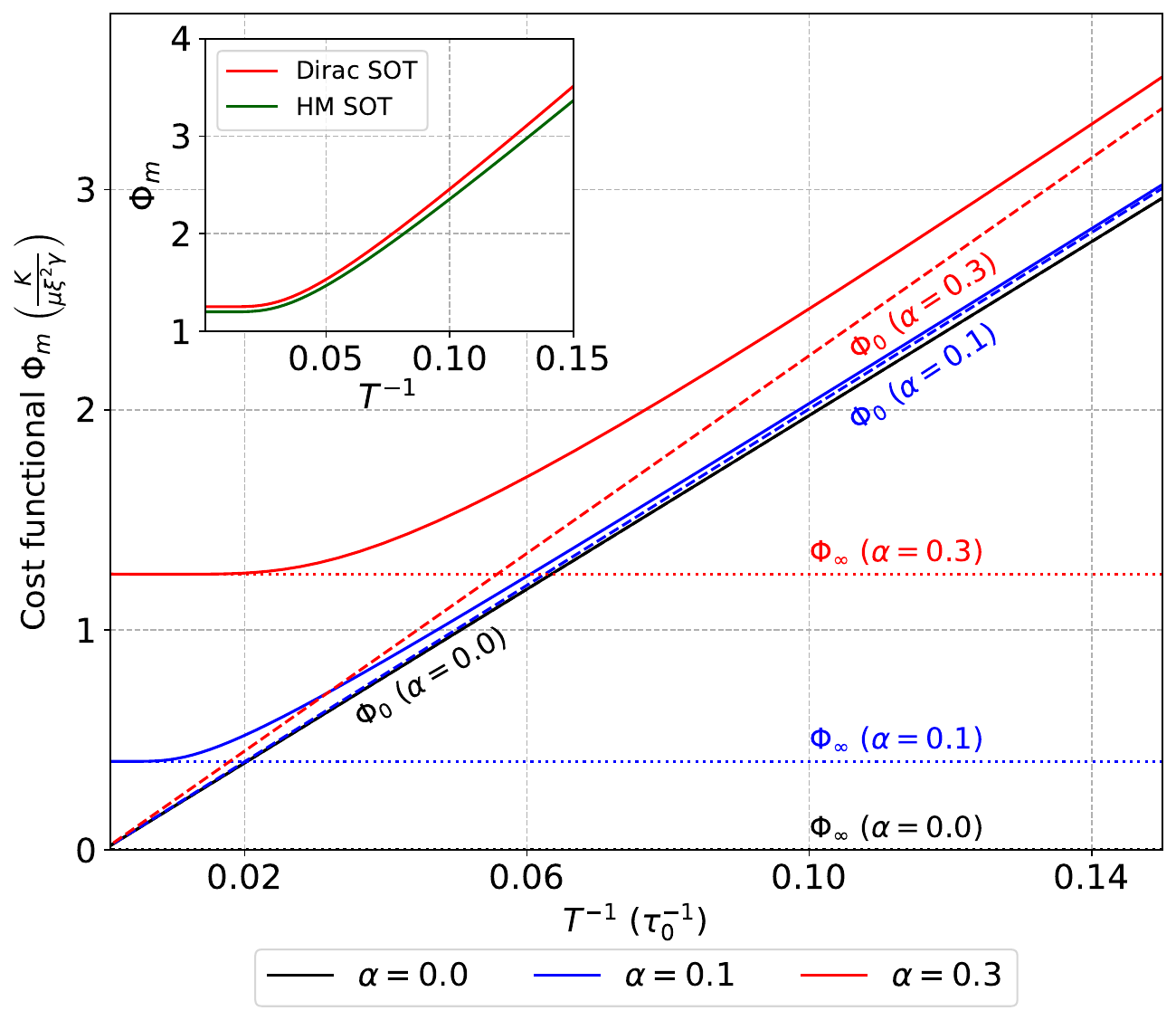}    
\vspace{-4mm}
\caption{(Color online) Calculated minimum energy cost $\Phi_m$ of the magnetization switching induced by the Dirac SOT as a function of $1/T$ for various values of the damping parameter as indicated in the legend. Dotted and dashed lines show the long  and short switching time asymptotics, respectively. \textit{Inset:} comparison of $\Phi_m$ for the Dirac and HM spin-orbit torques for $\alpha=0.3$. In all cases, $\beta$ is chosen to represent the lowest switching cost, i.e., $\beta=0$ for the Dirac SOT and $\beta=-\arctan{\alpha}$ for the HM SOT.}
\label{fig:cost-inverse-time}
\end{figure}

$\Phi_m$ as a function of the inverse switching time is shown in Fig. \ref{fig:cost-inverse-time}, where we assume the optimal values of $\beta$ ($\beta=-\arctan\alpha$ for FM/HM~\cite{vlasov2022} and $\beta=0$ for FM/TI). As can be seen in the inset, the energy efficiencies of switching are very similar for both SOT types. This can be explained by the fact that the optimal switching relies mostly on the FL torque in both cases, and its form is the same in the models describing FM/TI [Eq.\ (\ref{eq:eom})] and FM/HM \cite{vlasov2022} systems. However, the former presents only a slightly higher cost for switching because, in an FM/HM configuration, the DL torque can compensate for relaxation, whereas in an FM/TI system with Dirac SOT it cannot. For typical values of $\alpha$, and even larger (see Appendix \ref{sec:realistic-parameters}), this difference in switching cost between the Dirac and HM SOT is small ($\approx 4\%$ when $\alpha=0.3$). At the same time, topological insulators have the advantage of channeling the whole current through the surface instead of in the bulk, while electrons present a spin orientation fixed by their direction of motion, increasing the spin-torque efficiency, as demonstrated by experiments \cite{Mellnik2014,Wang2015}. 

Away from the optimal regime of $\beta=0$, i.e. when $\xi_D$ becomes finite in Dirac-SOT systems (expected to be common experimental situations), an optimal protocol still exists; in this case, both $\Phi_m$ and $\vec{j}_m$ can be directly calculated with the supporting code provided in the SM.

\begin{figure}[h]
\centering
\includegraphics[width=220pt]{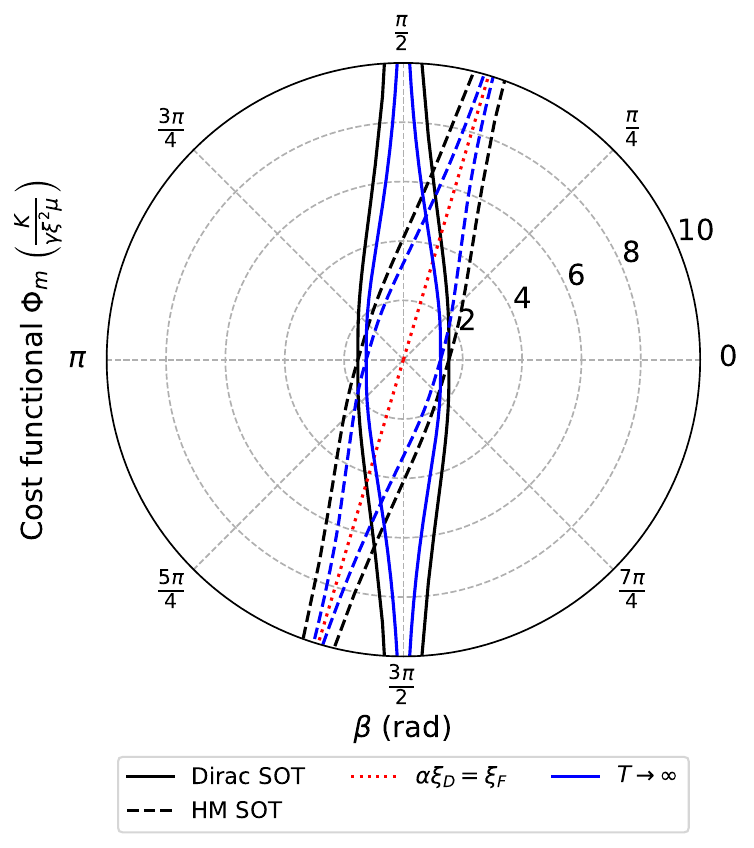}    
\vspace{-4mm}
\caption{(Color online) Calculated minimum energy cost of magnetization switching $\Phi_m$ as a function of $\beta$ for $\alpha=0.3$ and $T=20\tau_0$ (black lines) and $T\rightarrow\infty$ (blue lines), considering two different spin-orbit torques: Dirac (solid lines), and HM SOT (dashed lines). The red dotted line represents the DL and FL parameters ratio for which no switching occurs for the HM SOT.}
\label{fig:cost-beta}
\end{figure}

There is a strong dependence of $\Phi_m$ on the $\beta$ parameter, as shown in Fig. \ref{fig:cost-beta}. As predicted from Eq. (\ref{eq:current_amplitude_final}), we obtain the lowest energy cost for $\beta=0$ and $\beta=\pi$, while for $\beta=\pi/2$ and $\beta=3\pi/2$ the cost diverges to infinity, indicating that switching is not possible using just SOT. Unlike the  
HM SOT, the optimal value of $\beta$ and the value at which switching becomes impossible are independent of $\alpha$ for TI SOT. Nevertheless, both cases exhibit very similar dependencies of the switching cost on $\beta$. However, they are offset by $\Delta\beta = \arctan\alpha$ relative to one another, due to the differing geometrical forms of the DL torque component. 

\subsection{Tunability of SOT  coefficients}\label{sec:tunability}

Having analyzed the lowest energy cost and corresponding currents $\vec{j}_m(t)$ and identified the optimal ratio of the SOT coefficients $\beta=0$, we now link $\beta$ to the physical parameters of an FM/TI system and briefly explore their potential tunability.

The coefficients of the SOT acting on a magnet as a result of current flowing at the surface of a TI are given by the following equations \cite{ndiaye17:014408}:
\begin{equation}
\xi_F \approx e\rho \frac{J_{xc}\tau \varepsilon_F}{\pi \hbar^2 v_F \gamma}\,, \,\textnormal{and }
\xi_D \approx -e\rho\frac{2J_{xc}^2}{\pi \hbar v_F \varepsilon_F \gamma}\,,
 \label{eq:torqueTI}
\end{equation}

\noindent where $e>0$ is the elementary charge, $\hbar$ the reduced Planck constant, $J_{xc}$ the exchange coupling between the FM and the TI, $\tau$ the charge carrier scattering time, and $\varepsilon_F$ and $v_F$ are the Fermi energy and Fermi velocity, respectively.
The exchange parameter describes the coupling between the magnetization of the FM and the out-of-equilibrium Dirac state spin density on the surface of the adjacent TI. This is an interface property that depends on the specific materials used, but it can also be tuned by interface engineering \cite{Luo085431:2013,Jiang15:5835}. When $J_\mathrm{xc}/\varepsilon_F \ll 1$, both $\xi_F$ and $\xi_D$ become approximately independent of the magnetization state~\cite{ndiaye17:014408}; this regime can be achieved in real systems (see discussion in Appendix \ref{sec:realistic-parameters}). In turn, the sample composition and quality determine both $\tau$ and $v_F$. Here, we assume an electric field $\vec{E}=\rho\vec{j}$, where $\rho$ is some relevant surface resistivity. Interestingly, in a normal 2D system with Rashba spin-orbit coupling (SOC), the SOT takes the same form as in Eq.\ (\ref{eq:torqueTI}) when $J_{xc} \ll \alpha_R k_F$, with $\alpha_R$ being the Rashba coupling parameter~\cite{li15:134402}.

Comparing Eqs. (\ref{eq:TIcomponents}) and (\ref{eq:torqueTI}) reveals that the $\beta$ parameter is connected to the microscopic quantities through the following equation: 
\begin{equation}
    \tan (\beta)=\frac{-2J_{xc}\hbar}{\varepsilon_F^2 \tau},
    \label{eq:tanBeta}
\end{equation}
and the magnitude of the total SOT is given by
\begin{equation}
    \xi=\left |\frac{J_{xc}}{\pi \hbar v_F} \frac{e\rho}{\gamma}\right |\sqrt{\left ( \frac{\varepsilon_F \tau}{\hbar} \right )^2+\left ( \frac{ 2 J_{xc}}{\varepsilon_F} \right )^2}.
\end{equation}
The $\beta$ parameter does not explicitly depend on $v_F$, but $\xi$ is, however, inversely proportional to $v_F$.  By applying strain \cite{Aramberri2017}, or external gate voltages \cite{DiazFernandez2017}, the value of the Fermi velocity can be modified, and in some cases made asymmetric \cite{Aramberri2017}.   The torque parameters in Eq. (\ref{eq:torqueTI}) only contain the Fermi surface contribution, relevant when the Fermi energy is in the gap.  For $\varepsilon_F$  closer to bulk bands, additional terms appear due to, e.g., hexagonal warping -- which can be thought of as the counterpart of the cubic Dresselhaus coupling -- \cite{Fu2009,Farokhnezhad2024}, and spin-transfer torque due to bulk states \cite{Ghosh2018,Cullen2023,Farokhnezhad2024}, but these are beyond the scope of this article.

The $\beta$ parameter depends on $J_\mathrm{xc}, \tau$, and $\varepsilon_F$, as shown in Eq.\ (\ref{eq:tanBeta}).  The exchange coupling and electron scattering time can be tuned by choosing different materials or via interface engineering \cite{Luo085431:2013,Jiang15:5835}.  In addition, $\varepsilon_F$ can readily be tuned in TIs by gate voltages, through electron density changes, or even light pulses \cite{Taupin2023}; in HM systems, however, the Fermi energy is fixed.

In case of FM/HM systems, the ratio between the torque coefficients can be obtained from the model of an FM two-dimensional electron gas with Rashba SOC. This approach is commonly used as an archetypal free-electron model for SOT in ultrathin FM layers sandwiched between two distinct nonmagnetic materials \cite{li15:134402}. Within this model, the relationship between the torque coefficients in the context of the $\beta$-parametrization [see Eq. (\ref{eq:TIcomponents})] is determined by the following equation: 
\begin{equation}
    \tan (\beta^{\textnormal{HM}})=\frac{J_{xc} \hbar^3}{2\varepsilon_F \tau \alpha_R^2 m^{*}},
    \label{eq:tanBetaHM}
\end{equation}

\noindent where $m^{*}$ is the effective charge carrier mass \footnote{The difference in the parametric dependence in Eqs.\ (\ref{eq:tanBeta}) and (\ref{eq:tanBetaHM}) is due to the distinct density of states in TI and HM systems. In particular, $D_\mathrm{TI}(\varepsilon_F)=\frac{1}{2\pi}\frac{\varepsilon_F}{\hbar^2 v_F^2}$ for TIs and $D_\mathrm{HM}(\varepsilon_F)=\frac{m^*}{\pi \hbar^2} $ for Rashba HM systems}. The high electron densities in FM/HM multilayers make it difficult to tune either $\alpha_R$ \cite{Kang2021} or $\varepsilon_F$. 

\subsection{A simplified switching protocol}
\label{sec:simplified-switching-protocol}

\begin{figure}
    \centering
    \includegraphics[width=1\linewidth]{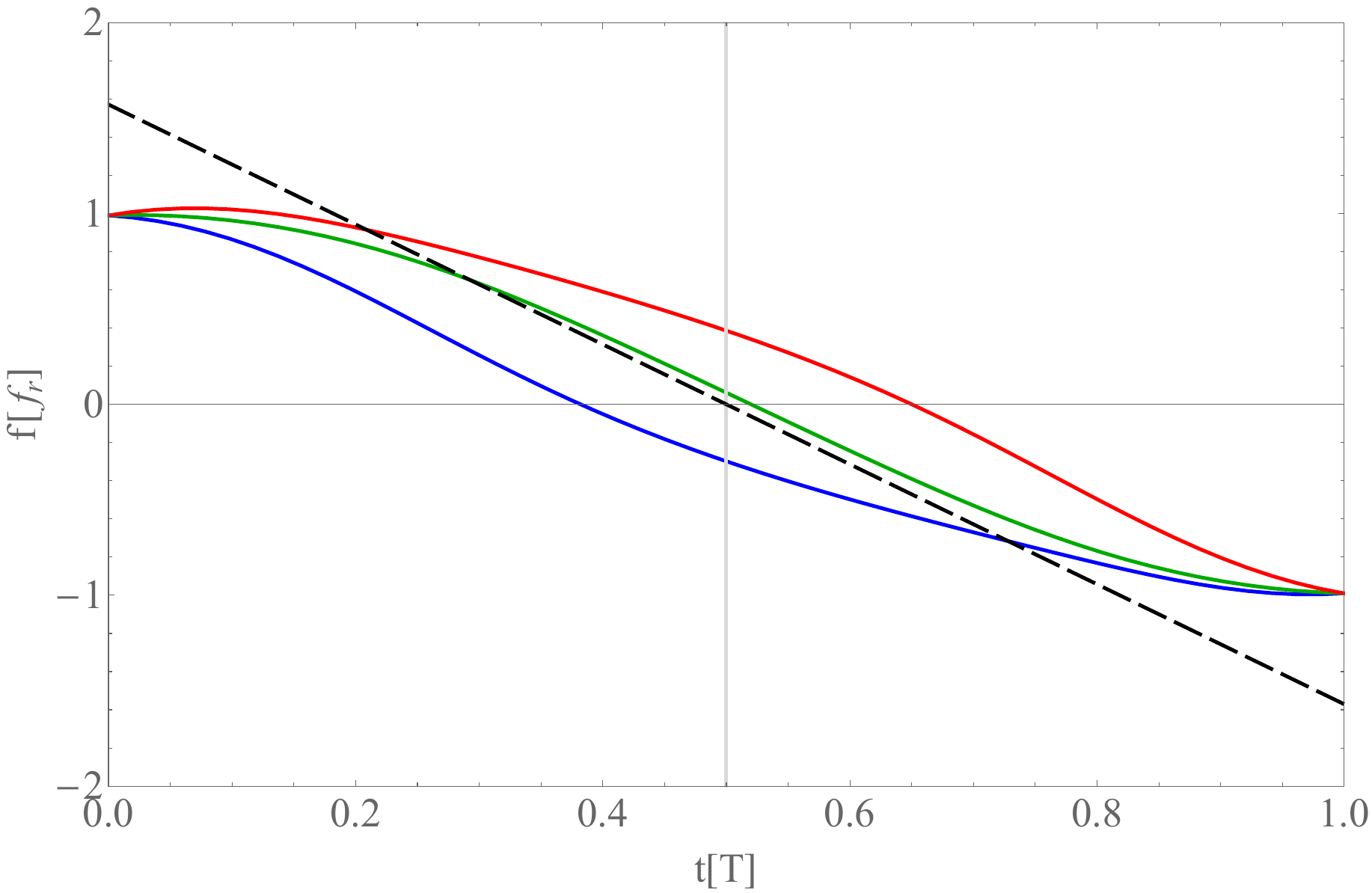}
    \caption{Calculated frequency of the optimal switching current as a function of time for $\beta = -1,\ 0,\ 1$ (blue, green, and red curves, respectively) for $\alpha = 0.1$ and $T = 20\tau_0$. The dashed black line represents the frequency $f(t)$ of the simplified pulse given by Eq. (\ref{eq:frequency}).}
    \label{fig:frequency}
\end{figure}

In general, the optimal switching current exhibits a complex, non-linear modulation of both frequency and amplitude (see Figs. \ref{fig:evolution-optimal-current} and \ref{fig:frequency}). This raises the question of whether a simpler, more practical switching protocol can be developed -- one that is easier to reproduce experimentally while maintaining energy efficiency close to that of the optimal protocol. In this context, we first point out that the optimal switching current maintains an almost constant amplitude in the low damping regime (see Fig. \ref{fig:evolution-optimal-current}). Simultaneously, when the DL torque vanishes and $T=T_{\textnormal{opt}}\gtrsim10\tau_0$, the time dependence of the frequency is nearly linear over most of the time domain (see Fig. \ref{fig:frequency}). Motivated by these observations, we focus on a protocol that involves the following down-chirped rotating current: 
\begin{equation}
\label{eq:simplified-pulse}
\begin{split}
    \vec{j}_{s}(t) &= A\left[\cos\vartheta(t), \sin\vartheta(t), 0\right]
\end{split}
\end{equation}

\noindent where $A$ is a time independent amplitude and $\vartheta(t)=\omega_0 \left(t - t^2/T\right)$, with  $\omega_0$ being  the initial angular velocity. Its rotation frequency is thus defined as

\begin{equation}
\label{eq:frequency}
f(t)=\frac{1}{2\pi}\frac{d\vartheta}{dt}=\frac{\omega_0}{2\pi}\left(1-\frac{2t}{T}\right).
\end{equation}

Although chirped signals have not yet been applied to field-free magnetization switching via current pulses, they have been successfully utilized in microwave photonics \cite{Li2014,Raghuwanshi2017}. These well-established techniques could provide valuable 
insights for generating current pulses similar to those  
 described by Eq. (\ref{eq:simplified-pulse}). 

The initial frequency was chosen so that the quantity $d^2\vartheta/dt^2$ roughly coincides with that of the optimal pulse (namely $d^2\psi_m/dt^2$) near the energy barrier defined by the anisotropy -- i.e., in the vicinity of $t = T/2$. Figure \ref{fig:frequency} illustrates an example of the computed frequency, in units of the resonant frequency $f_r\equiv[2\pi\tau_0(1+\alpha^2)]^{-1}$, of the optimal pulses for distinct $\beta$ values and $T=20\tau_0$, together with that proposed for the simplified pulse. 
Here, we clearly see that, when $\beta$ strongly deviates from zero (e.g., $\beta=\pm1$, shown by the red and blue curves), the optimal frequency becomes more non-linear over the time domain, and no longer aligns with the proposed rotation frequency [Eq. (\ref{eq:frequency})]. Moreover, the cost functional for Eq.\ (\ref{eq:simplified-pulse}) reduces to $\Phi = A^2T$, which defines a lower bound on the amplitude, $A_m = \sqrt{\Phi_m/T}$, determined by the minimum energy cost $\Phi_m$. 

\begin{figure*}[htb!]
\centering
\includegraphics[width=1.5\columnwidth]{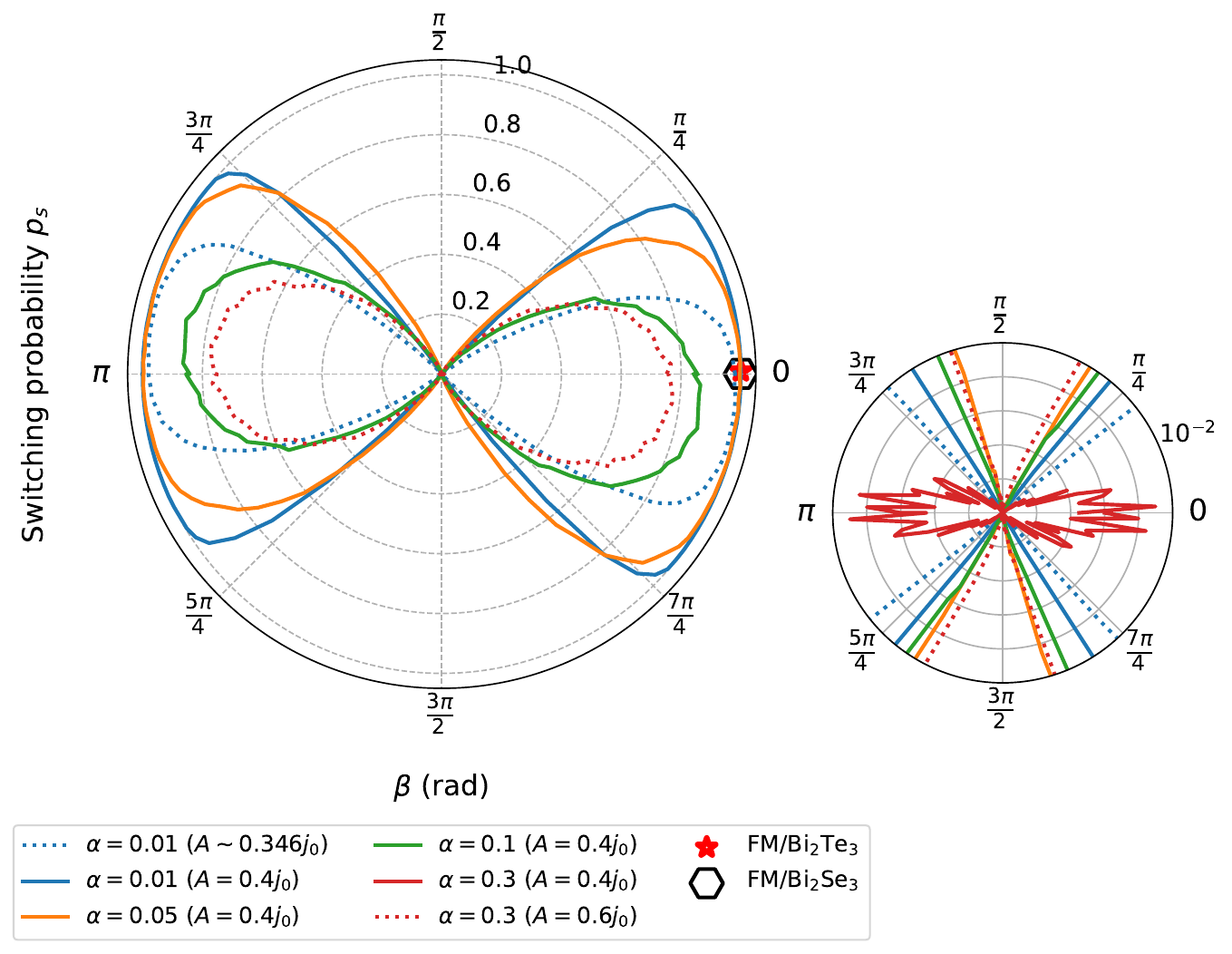}    
\vspace{-4mm}
\caption{(Color online) Calculated switching probability $p_s$ as a function of $\beta$ for $T=20\tau_0$ and selected Gilbert damping values: for $A=0.4j_0$ (solid lines); for different choices of amplitude $A$ (dashed lines). The red star and open black hexagon mark realistic $\beta$ values for FM/TI systems, as discussed in Appendix \ref{sec:realistic-parameters}. \textit{Inset}: a zoom on the range $p_s\in[0,10^{-2}]$, in which the results for $\alpha=0.3$ and amplitude $A=0.4j_0$ become visible.}
\label{fig:switchin-prob}
\end{figure*}

To test the magnetization switching using the simplified pulse, we performed direct time integration of the LLG equation (Eq. \ref{eq:eom}) including a stochastic term to simulate the thermal noise, and considering $\vec{j}(t)=\vec{j}_{s}(t)$. In this case, the choice of temperature matches the required stability factor for a memory element, defined as the ratio between the energy barrier and the thermal energy, of 60 \cite{vlasov2022} \footnote{The stability factor assumed here is a conservative lower threshold for a 10-year data retention time. In practice, the commonly required factor for MRAM systems is $\approx70$ or more \cite{Chun2013}.}. Additionally, the amplitude was set to be close to $A_m$. 
The simulations consisted of three steps: (\textit{i}) the initial phase where no pulse is applied in order to establish the thermal equilibrium; (\textit{ii}) the measurement phase including thermal fluctuations, where the simplified current pulse is applied over the time interval $T$; and (\textit{iii}) the final thermal equilibration with no applied current. After step (\textit{iii}), the switching event is considered successful if $s_z<-0.5$, and the switching probability $p_s$ is then defined as \cite{vlasov2022}

\begin{equation}
\label{eq:switching-probability}
p_s = \frac{N_s}{N},
\end{equation}

\noindent where $N_s$ is the number of successful switching events, and $N$ is the total number of simulations. We choose $N=2000$, which ensures that the shape of the resulting $p_s$ curve as a function of $\beta$ is properly converged. In Figure \ref{fig:switchin-prob} we show the calculated $p_s(\beta)$ curves for $T=20\tau_0$,  $A=0.4j_0$, and 
several values of the damping parameter, namely $\alpha=\{0.01,0.05,0.1,0.3\}$, compatible with those found in FM/TI heterostructures (see Appendix \ref{sec:realistic-parameters}). As can be seen, the small realistic $\beta$ values for $\alpha=0.01$ and $\alpha=0.05$ correspond to the plateau region with $p_s\approx1$. The switching probability decreases with $\alpha$. Still, $p_s\gtrsim0.8$ for $\alpha=0.1$. We note that for $\alpha = 0.01$, selecting an amplitude $A$ just $10\%$ greater than $A_m$ (i.e., $A\approx0.346j_0$ in this case) results in a switching probability close to $100\%$ near $\beta = 0$ \footnote{Using the values given in Ref. \cite{Ghosh2018}, we find $p_s\approx1$ for $\alpha=0.01$ and $A=0.4j_0$.}. This corroborates with the idea that $\vec{j}_{s}$ serves as an excellent model for the optimal pulse when $\beta$ is sufficiently small.

For the highest Gilbert damping analyzed, $\alpha=0.3$, the amplitude has to be increased to at least $A=0.6j_0$ to reach $p_s\approx0.8$ -- or $\sim50\%$ larger than the value $A_m$ for that value of $\alpha$. Nevertheless, switching can be robustly achieved in various realistic setups (specifically considering the typical small $\alpha$ values for FM insulators when compared to FM metals -- see Appendix \ref{sec:realistic-parameters}), while still ensuring an energy cost that remains close to the minimum value $\Phi_m$. In this sense, we also note that no fine-tuning of pulse duration is required.

The simplified pulse proposed in Eq. (\ref{eq:simplified-pulse}) is effective for small $\beta$, as observed in certain FM/TI systems. For materials where $\beta$ deviates strongly from zero, achieving reliable, field-free switching may require alternative pulse designs. In such cases, the pulse in Eq. (\ref{eq:simplified-pulse}) cannot ensure reliable switching on the timescale of moment oscillations (see Fig.~\ref{fig:switchin-prob}). 
For Dirac-SOT systems with large $\beta$, the realization of efficient magnetization switching via down-chirped rotating current requires the $\beta$ value to be reduced, e.g., via engineering of $\varepsilon_F$ and $\tau$. Although chirped pulses can induce slower reversals via the autoresonance mechanism \cite{Go2017}, this approach is likely sensitive to thermal fluctuations, which may disrupt the frequency locking between the oscillator and the excitation.

\section{Conclusions and outlook} 

We investigated the energy-efficient switching process in a single macrospin controlled by external currents when subject to the Dirac SOT that is predicted to arise in FM/TI systems \cite{ndiaye17:014408}. We used the optimal control theory to obtain the energy-efficiency limits of Dirac SOT-induced magnetization switching in a PMA nanoelement  at zero applied magnetic field and derived the optimal time-dependent switching current. We established that the FL component of the Dirac SOT is only responsible for switching and the energy cost is lowest for vanishing DL torque. This represents a fundamental difference to the HM SOT \cite{manchon2019,vlasov2022}, arising typically in FM/HM systems, in which the DL torque may compensate for the relaxation and reduce the overall energy cost of the switching. 

The obtained minimal switching cost is comparable 
to that obtained for the HM SOT~\cite{vlasov2022}. Results for the respective types of SOTs coincide in the undamped case and the difference between them increases with increasing Gilbert damping. For the damping parameter values expected in the PMA nanoelement on a TI substrate, the energy cost difference is estimated to be less than $\approx 1\%$ under the assumption of the same model parameters and respective optimal ratio of the torque coefficients for both SOT models. This shows that magnetic systems exhibiting Dirac SOT are not only viable candidates for memory elements, but present several advantages over FM/HM heterostructures, such as: (\textit{i}) the tunability of the physical parameters, (\textit{ii}) spin-momentum locking of electron states (which results in a large spin-charge conversion and improved SOT efficiency), and (\textit{iii}) currents necessary for switching that are one order of magnitude lower, as realized in room temperature experiments \cite{Wang2017,Khang2018}, where the SOT efficiency reflects in a higher $\xi$ parameter, and, thus, in a lower energy cost of switching. 
 
 Our analysis naturally has the same limitations as the model developed by Ndiaye \textit{et al.} \cite{ndiaye17:014408}, primarily stemming from the assumption that the TI surface states remain unaffected in the presence of the FM layer, and that transport is strictly two-dimensional, occurring solely on the  surface of the TI. When these assumptions are not met, the introduction of an FM layer can lead to more complex scenarios. For instance, Mahfouzi \textit{et al.} \cite{Mahfouzi2016} report a non-trivial dependence of the SOT coefficients on the magnetization direction, whose analysis in the context of the optimal control theory is a subject of future research. 

Finally, we propose a simplified protocol based on a down-chirped rotating current pulse, which can be more accessible to experiments. This protocol was tested through simulations of the stochastic LLG equation, and 
the results demonstrate its efficiency and robustness across a range of realistic $\beta$ values. These theoretical predictions motivate experimental validation.

\begin{acknowledgments} 
The authors would like to thank T. Sigurj\'onsd\'ottir for helpful discussions and useful comments. 
This work was supported by the Icelandic Research Fund (Grant No. 217750 and No. 217813), the University of Iceland Research Fund (Grant No. 15673), the Faculty of Technology and the Department of Physics and Electrical Engineering at Linnaeus University (Sweden), the Swedish Research Council (Grants No. 2020-05110 and No. 2021-046229), the Russian Science Foundation (Grant no. 23-72-10028), the Carl Tryggers Foundation (CTS20:71), and the Crafoord Foundation (grant No. 20231063). The computations/data handling were enabled by resources provided by the National Academic Infrastructure for Supercomputing in Sweden (NAISS), partially funded by the Swedish Research Council through grant agreement No. 2022-06725.
\end{acknowledgments}

\appendix

\section{Realistic $\alpha$ and $\beta$ for FM/TI systems}
\label{sec:realistic-parameters}

In Section \ref{sec:tunability}, we express the $\beta$ parameter describing the ratio of the SOT coefficients in terms of material-specific microscopic quantities [see Eq.~(\ref{eq:tanBeta})]. 
Typical values for $J_{xc}$ are around $0.05$ eV \cite{Li2019}, or even of the order of $\sim10^{-1}$ eV if the exchange is considered to be between charge carriers and magnetic impurities \cite{Checkelsky2012,Liu2009}, although it can be as low as $J_{xc}\approx0.006$ eV in the Ni$_{81}$Fe$_{19}$/Bi$_{2}$Te$_{3}$ junction \cite{Shiomi2014}. For ferromagnetic insulators, the $J_{xc}$ values are expected to be smaller than in metallic systems, as the $d$ or $f$ states are more localized, and the exchange interaction is typically limited to nearest neighbors \cite{Gorbatov2021}; however, proximity to the TI can enhance these exchange interactions \cite{Kim2017}. Therefore, to roughly estimate the absolute maximum value of $\beta$ given by Eq.\ (\ref{eq:tanBeta}), $\left|\beta_{\textnormal{max}}\right|$, we consider 
$J_{xc}=0.1$ eV. In Table \ref{tab:parameters-tis}, we present some experimental and theoretical values collected from the literature. Obviously, these parameters can vary due to several factors, such as temperature, the thickness of the FM layer or the TI, impurity types/concentrations, among others; in fact, the ability to tune these parameters is one of the key advantages of magnetization switching via SOT in FM/TI heterostructures, as discussed in Section \ref{sec:tunability}. Nevertheless, the values provided here should offer a useful insights into realistic estimates of $\beta$.

As the results have been shown to be sensitive to the Gilbert damping $\alpha$, here we also present a collection of values available from previous research (Table \ref{tab:damping-fm-ti}). Even though proximity effects with TIs enhance $\alpha$, FM insulators in their bare form generally exhibit lower damping than FM metals. Therefore, the overall (enhanced) damping value, when coupled with TIs, still tends to remain relatively lower than that in such metals.
 
\begin{table*}[!htb]
\caption{Parameters $\tau$ and $\varepsilon_F$ of some known topological insulators. From those values, $\left|\beta_{\textnormal{max}}\right|$ is calculated according to Eq.~(\ref{eq:tanBeta}).}
\begin{center}
\label{tab:parameters-tis}
\begin{tabular}{c c c c}
\hline 
\hline
& \\[-2.5mm] 
\textbf{Compound} & $\tau$ (ps) & $\varepsilon_F$ (eV) & $\left|\beta_{\textnormal{max}}\right|$ (rad)\\ 
& \\[-3mm] \hline
& \\[-3mm] 	    
\multirow{2}{*}{Bi$_2$Te$_3$} & $>0.25$ \cite{Plank2018} & \multirow{2}{*}{$\sim0.37$\footnotemark[3] \cite{Scholz2013}} & \multirow{2}{*}{$\sim4\times10^{-3}$-$10^{-2}$} \\
& $\sim0.1$-$0.25$\footnotemark[1] \cite{Chapler2014} & & \\ \hline
\multirow{3}{*}{Bi$_2$Se$_3$} & $\sim0.45$-$1$\footnotemark[2] \cite{Wu2013} & \multirow{3}{*}{$\sim0.5$\footnotemark[4] \cite{Zhang2016}} & \multirow{3}{*}{$\sim5\times10^{-4}$-$10^{-3}$} \\
& $\sim0.4$-$0.8$\footnotemark[2] \cite{Sim2014} & & \\
& \multirow{1}{*}{$\sim0.8$ \cite{Autore2017}} & & \\
\\[-3mm] \hline \hline
\end{tabular}
\end{center}	
\footnotetext[1]{The lowest (highest) scattering time refers to the 4.5\% Mn-doped (pristine) Bi$_2$Te$_3$ epitaxial thin films.}
\footnotetext[2]{The interval represents different thicknesses of the topological insulator films, with the lowest (highest) $\tau$ values corresponding to the largest (smallest) thickness.}
\footnotetext[3]{Fermi level for a monolayer of Fe deposited in Bi$_2$Te$_3$, with respect to the Dirac point. For the pristine sample (i.e., without Fe deposition), $\varepsilon_F\approx0.22$ eV \cite{Scholz2013}.}
\footnotetext[4]{Calculated via first principles for the system Bi$_2$Se$_3$/Ni(Co) consisting of 6 monolayers of the magnetic material. $\varepsilon_F$ position with respect to the Dirac cone in a stand-alone Bi$_2$Se$_3$(0001) surface.}
\end{table*}

\begin{table*}[!htb]
\caption{Gilbert damping values ($\alpha$) for FM/TI heterostructures.}
\begin{center}
\label{tab:damping-fm-ti}
\begin{tabular}{cc c c}
\hline 
\hline
& & \\[-2.5mm] 
 & \textbf{System} & $\alpha$ & \textbf{Origin} \\ 
& & \\[-3mm] \hline
& & \\[-3mm]
\multirow{6}{*}{\parbox[t]{2cm}{\centering \textbf{Ferromagnetic} \\ \textbf{insulators}}}
& {Y$_3$Fe$_5$O$_{12}$/Bi$_2$Te$_3$  (expt.) \cite{Will-Cole2024}} & \multirow{2}{*}{$\sim0.01$\footnotemark[1]} & Intrinsic/spin pumping/ \\
& & & proximity effect \\
& {Y$_3$Fe$_5$O$_{12}$/Bi$_2$Se$_3$  (expt.) \cite{Liu2020}} & \multirow{2}{*}{$\sim0.004$-$0.05$\footnotemark[2]} & Intrinsic/spin pumping/ \\
& & & proximity effect \\ 
& {Y$_3$Fe$_5$O$_{12}$/(Bi$_x$Sb$_{1-x}$)$_2$Se$_3$  (expt.) \cite{Tang2018}} & \multirow{2}{*}{$\sim0.0045$-$0.022$\footnotemark[3]} & Intrinsic/spin pumping/ \\
& & & proximity effect \\ \hline
\multirow{14}{*}{\parbox[t]{2cm}{\centering \textbf{Ferromagnetic} \\ \textbf{metals}}} & Fe monolayer/Bi$_2$Se$_3$ (theory) \cite{Hou2019} & $\sim0.03$-$0.05$\footnotemark[4] & Intrinsic (spin-orbit) \\
& Co monolayer/Bi$_2$Se$_3$ (theory) \cite{Hou2019} & $\sim0.025$-$0.03$\footnotemark[4] & Intrinsic (spin-orbit) \\
& Ni monolayer/Bi$_2$Se$_3$ (theory) \cite{Hou2019} & $\sim0.1$-$0.25$\footnotemark[4] & Intrinsic (spin-orbit) \\
& \multirow{2}{*}{Co$_{80}$Fe$_{20}$/Bi$_2$Te$_3$ (expt.) \cite{Wu2021}} & \multirow{2}{*}{$0.023$\footnotemark[5]} & Intrinsic/spin pumping/ \\
& & & proximity effect \\
& \multirow{2}{*}{Co$_{80}$Fe$_{20}$/Sb$_2$Te$_3$ (expt.) \cite{Wu2021}} & \multirow{2}{*}{$0.025$\footnotemark[5]} & Intrinsic/spin pumping/ \\
& & & proximity effect \\
& \multirow{2}{*}{Co$_{80}$Fe$_{20}$/(Cr$_{0.2}$Bi$_{0.24}$Sb$_{0.56}$)$_2$Te$_3$ (expt.) \cite{Wu2021}} & \multirow{2}{*}{$0.015$\footnotemark[5]} & Intrinsic/spin pumping/ \\
& & & proximity effect \\
& \multirow{2}{*}{Ni$_{80}$Fe$_{20}$/Bi$_2$Te$_3$  (expt.) \cite{Bhattacharjee2022}} & \multirow{2}{*}{$0.0123$\footnotemark[6]} & Intrinsic/spin pumping/ \\
& & & proximity effect \\
& {CoFeB/Bi$_2$Te$_3$  (expt.) \cite{Bhattacharjee2022}} & \multirow{2}{*}{$0.283$\footnotemark[6]} & Intrinsic/spin pumping/ \\
& & & proximity effect \\
& {Ni$_{80}$Fe$_{20}$/Sb$_2$Te$_3$  (expt.) \cite{Will-Cole2023}} & \multirow{2}{*}{$0.178$\footnotemark[7]} & Intrinsic/spin pumping/ \\
& & & proximity effect
\\\hline\hline
\end{tabular}
\end{center}	
\footnotetext[1]{Measured at $T=10$ K for a 25-nm-thick Bi$_2$Te$_3$ grown by sputtering on a 50-nm-thick liquid phase epitaxy grown YIG (Y$_3$Fe$_5$O$_{12}$) on a Gd$_3$Ga$_5$O$_{12}$ host.}
\footnotetext[2]{The lowest value corresponds to room-temperature ferromagnetic resonance (FMR) measurements in a 8-nm-thick Bi$_2$Se$_3$ film deposited on a 15-nm-thick YIG film. In turn, the highest value corresponds to a temperature of $T=50$ K.}
\footnotetext[3]{Measurements at $T=300$ K on $\sim$5-nm-thick topological insulator deposited on 10-nm-thick YIG films. The highest value corresponds to a Bi concentration of $x=0.32$.  Damping value for the bare YIG films: $1.5\times10^{-3}$.}
\footnotetext[4]{For the broadening parameters $\Gamma=26$ and 100 meV.}
\footnotetext[5]{Thickness of 6 nm for the TI and 8 nm for the ferromagnetic layer. Measurements performed at $T=80$ K.}
\footnotetext[6]{For the most structurally ordered sample. Thickness of $\sim30$ nm for Bi$_2$Te$_3$ and $\sim20$ nm for the ferromagnet (Ni$_{80}$Fe$_{20}$ or CoFeB). Measurements performed at $T=300$ K.}
\footnotetext[7]{Highly $c$-axis oriented. Thickness of $\sim10$ nm for Sb$_2$Te$_3$ and $\sim15$ nm for Permalloy (Ni$_{80}$Fe$_{20}$). Measurements performed at $T=300$ K.}
\end{table*}

\bibliography{refs}

\end{document}